 \documentclass [12pt,a4paper      ]{article}
\usepackage{times}

\DeclareFontFamily{OT1}{times}{}
\DeclareFontShape {OT1}{times}{m }{n }{ <-> ptmr }{}
\DeclareFontShape {OT1}{times}{bx}{n }{ <-> ptmb }{}
\DeclareFontShape {OT1}{times}{m }{it}{ <-> ptmri}{}
\DeclareFontShape {OT1}{times}{bx}{it}{ <-> ptmbi}{}
\usepackage{amsmath}
\usepackage{amsfonts}
\usepackage{amssymb}
\usepackage{latexsym}
\newcommand{\cl}{C \kern -0.1em \ell} 

\newcommand{\CON}{\overline}          

\newcommand{\VEC}{\vec{\kern +.1em[}} 
\newcommand{\TOR}{\vec{\kern +.2em]}} 
\newcommand{\BRA}{\langle\kern -.2em\langle} 
\newcommand{\KET}{\rangle\kern -.2em\rangle} 

\newcommand{\DUA}{\widetilde}         

\numberwithin{equation}{section}
\begin{document}

\title{\bf\vspace{-2.5cm} On the Covariant Formulation of Dirac's
                          Equation\footnote{\emph{Editorial note:} Published
                          in Zeits.\ f.\ Phys.\ {\bf 57} (1929) 474--483,
                          reprinted and translated in~\cite{LAN1929C}.
                          This is Nb.~2 in a series of four papers
                          on relativistic quantum mechanics
                          \cite{LAN1929B,LAN1929C,LAN1929D,LAN1930A}
                          which are extensively discussed
                          in a commentary by Andre Gsponer
                          and Jean-Pierre Hurni \cite{GSPON1998B}.
                          Initial translation by J\'osef Illy and 
                          Judith Konst\'ag Mask\'o. Final translation
                          and editorial notes by Andre Gsponer.}}

\author{By Cornel Lanczos in Berlin\\ (Received on August 3, 1929)}

\date{Version ISRI-04-11.3 ~~ \today}

\maketitle

\begin{abstract}

As a continuation of previous investigations, the formalism used there is extended to the case when an external electric field is present and the covariant formulation is performed again. The equation system obtained allows no restriction in the manifold of the quantities if an undesired overdetermination is to be avoided. The combined appearance of the dual formations also indicates an improbable internal properly of the system and suggests that the underlying Dirac equation needs a modification. (\emph{Editorial note:} In this paper Lanczos continues to discuss his ``fundamental equation,'' from which two Dirac fields forming an isospin doublet, or two spin 1 fields of opposite parities, derive.)

\end{abstract}

In a previous paper\footnote{Zeits.\ f.\ Phys.\ {\bf 57}, 447, 1929, \text{subsequently quoted as loc.\ cit.\ (\emph{Editorial note:} Ref.~\cite{LAN1929B})} } we showed that Dirac's system for a free electron is closely related to the Hamiltonian quaternion operator. On the basis of the close relations of this operator to the tensor analytical formulation of the four-dimensional space, we arrived at a complete tensor analytical description. Operating with the imaginary unit proves to be a purely formal tool. In fact, it disappears from the resulting system, which then contains only real vector analytical quantities. In particular, there appears an antisymmetric tensor of the type of the electromagnetic field strength, as well as two vectors, which can be compared to the electric and magnetic currents of Maxwell's theory. Namely, the couplings between these quantities fully correspond to those in Maxwell's equations, the only exception being that there is a further coupling present which, together with the others, consequently determines the Schr\"odinger equation.\footnote{\emph{Editorial Note}: By Schr\"odinger equation Lanczos means the \emph{relativistic} Schr\"odinger wave equation which today is usually referred to as the Klein-Gordon equation.} Until now we have only dealt with the equation for a free electron, and now we want to extend the investigation to the case where an external electromagnetic field is also present.

   In the case of Dirac's equation, the rule for performing the extension is the following: The operation $\frac{\partial}{\partial x_i} $ is to be completed by adding $i \Phi_i$, where $\Phi_i$ means the external vector potential multiplied by $\frac{2 \pi}{h} e$:
\begin{equation*}\label{1}
\Phi_i = \frac{2 \pi}{h} e \varphi_i .
\tag{1}
\end{equation*}
It is obvious that this extension can be performed according to the same rule in our covariant system as well without disturbing the covariance. Namely, the components $\Phi_i$ transform in the same way as $\frac{\partial}{\partial x_i} $; and the appearance of $i$ causes no difficulty because arbitrary complex quantities may appear in our quaternion calculus although the real interpretation is naturally assured.

   Let us write down our basic equation (54) (loc. cit) once more and in a form which makes no use of the specific reality conditions of the actual space-time continuum and thereby ensures the actual covariance:
\begin{equation*}\label{2}
\left.
\begin{aligned}
\CON{\nabla} F & = \alpha G^* , \\
\nabla G^* & = \alpha F .
\end{aligned}
\right\} 
\tag{2}
\end{equation*}
Let us now apply the rule and replace the $\frac{\partial}{\partial x_i}$ operation by $\frac{\partial}{\partial x_i} + i \Phi_i $, obtaining thereby the following system:
\begin{equation*}\label{3}
\left.
\begin{aligned}
\CON{\nabla} F &+ i \CON{\Phi} F & & = \alpha G^*, \\
\nabla G^* &+ i \Phi G^* & & = \alpha F .
\end{aligned}
\right\}
\tag{3}
\end{equation*}
Here $\Phi $ is the quaternion representing the vector potential whose components are the $\Phi_i$. It is real in its spatial part and purely imaginary in its time part. Therefore the rule (just as for $\nabla$):
\begin{equation*}\label{4}
\Phi^* = - \CON{\Phi}
\tag{4}
\end{equation*}
is valid. Let us now write equation \eqref{3} once more by taking the conjugate complex in the second equation.  Then we obtain our basic equations with the desired result in the form:
\begin{equation*}\label{5}
\left.
\begin{aligned}
\CON{\nabla} F + i \CON{\Phi} F & = \alpha G^* , \\
\CON{\nabla} G - i \CON{\Phi} G & = - \alpha F^* .
\end{aligned}
\right\}
\tag{5}
\end{equation*}

   Before discussing the whole equation system, let us first consider Dirac's equation which must form a subsystem of the complete system. We saw [cf., equation (61), loc. cit.] that such a subsystem exists for the combination:
\begin{equation*}\label{6}
H = \frac{1}{2} [(F+G) + i (F-G) j_z ] ,
\tag{6}
\end{equation*}
just as it holds for the combination:
\begin{equation*}\label{6'}
H' = \frac{1}{2} [(G+F) + i (G-F) j_z ] .
\tag{6'}
\end{equation*}
By addition or subtraction of equations \eqref{5} we have:
\begin{equation*}\label{7}
\left.
\begin{aligned}
\CON{\nabla} (H + H') - \CON{\Phi} (H - H') j_z & = \alpha i (H - H')^* j_z , \\
\CON{\nabla} (H - H') - \CON{\Phi} (H + H') j_z & = \alpha i (H + H')^* j_z , 
\end{aligned}
\right\}
\tag{7}
\end{equation*}
and from this we obtain by addition or subtraction:
\begin{equation*}\label{8}
\left.
\begin{aligned}
\CON{\nabla} H  - \CON{\Phi} H  j_z & =   \alpha i H^* j_z , \\
\CON{\nabla} H' + \CON{\Phi} H' j_z & = - \alpha i H'{}^* j_z . 
\end{aligned}
\right\}
\tag{8}
\end{equation*}
These are two Dirac equations for $H$ or $H'$ alone,\footnote{\emph{Editorial note:} The correct interpretation of this pair as an isospin doublet was first given by Feza G\"ursey in 1957.  See section~7 in Ref.~\cite{GSPON1998B}.} which altogether are obviously equivalent to the initial system \eqref{5}.  It is easy to see that the introduction of the vector potential actually means replacing the operation $\frac{\partial}{\partial x_i}$ by  $\frac{\partial}{\partial x_i} + i \Phi_i$.  Namely, from assignment (60), loc. cit., it can immediately be seen that multiplication of $\psi$ by $i$ is equivalent to $H$ being multiplied by $-j_z$.  It is remarkable that the vector potential occurs with opposite sign in both equations, so the first equation contains the operation  $\frac{\partial}{\partial x_i} + i \Phi_i$ whereas the second contains the operation $\frac{\partial}{\partial x_i} - i \Phi_i$.

   It is interesting to analyze the transformation properties of the completed Dirac equation. Let us perform a Lorentz transformation once more [cf. equation (18), loc. cit.] by putting:
\begin{equation*}\label{9}
\left.
\begin{aligned}
\nabla ' & = p \nabla \CON{p}^* , \\
\Phi '   & = p \Phi   \CON{p}^* ,
\end{aligned}
\right\}
\tag{9}
\end{equation*}
and let the transformation of $H$ be again [cf., equation (85), ibid.]:
\begin{equation*}\label{10}
H ' = p H k .
\tag{10}
\end{equation*}
[This $H'$ has nothing to do with the $H'$ in equation \eqref{8}.] 

Now we have two conditions for quaternion $k$:
\begin{equation*}\label{11}
\left.
\begin{aligned}
k^* j_z & = j_z k , \\
k\,j_z & = j_z k . 
\end{aligned}
\right\}
\tag{11}
\end{equation*}
The second condition comes from the term with the vector potential and is responsible for $k$ being real:
\begin{equation*}\label{12}
k = k^* .
\tag{12}
\end{equation*}
Then, however, $k$ can have only one $j_z$ and one $j_l$ component. Thus, there remain only two degrees of freedom which can be reduced to a single one by normalizing the length. However, for the $\psi$ this means only one further phase transformation, which is also permitted in quantum mechanics.

   Accordingly, the group of transformations which we referred to in the previous paper (see Section 9) will fail, and thus the customary transformation theory of the $\psi$ quantities is valid. Also, the objections to the current vector of Dirac's theory cannot be maintained any longer, and, in fact, the covariance of the current vector is guaranteed.

   It seems all the more strange that the quaternion formed from $F$ and $G$ during the covariant extension, which corresponds to a divergence free current vector:
\begin{equation*}\label{13}
F \CON{F}^* + G \CON{G}^* ,
\tag{13}
\end{equation*}
does not represent any vector and has no vector analytical meaning at all. (It has a vector character only under purely spatial rotations.)\footnote{Remark during proofreading: The simple and fundamental meaning of this construction, namely that it represents the energy current which can be completed by the momentum current to form a tensor of second-order, was realized by the author only after the completion of this analysis. Cf., the paper ``The conservation laws in the field theoretical description of Dirac's theory'' to be published in the journal (\emph{Editorial note:} Ref.~\cite{LAN1929D}). }

   To deduce Schr\"odinger's wave equation, let us now apply the operation $\nabla$ to the first of equations \eqref{5}:
\begin{align}
\nabla \CON{\nabla} F + i \nabla ( \CON{\Phi} F) 
    &= \alpha \nabla G^* = \alpha ( \alpha F - i \Phi G^* ) \notag \\
    &= \alpha^2 F - i \Phi \CON{\nabla} F + \Phi \CON{\Phi} F . \label{14}\tag{14}
\end{align}
However,
\begin{equation*}\label{15}
\nabla ( \CON{ \Phi } F ) = (\nabla \CON{ \Phi }) F + \CON{ \Phi ( \CON{\nabla} } F) ,
\tag{15}
\end{equation*}
and if we take the right-hand side terms, then we shall realize that, except for the characteristic term $(\nabla \CON{\Phi} )F$ which describes the electron spin for Dirac's theory, all remaining operators obtain a scalar character. For obviously:
\begin{equation*}\label{16}
\Phi \CON{\nabla} + \CON{ \Phi \CON{\nabla} } = 2 \Phi_{\nu} \frac{\partial}{\partial x_{\nu}} ,
\tag{16}
\end{equation*}
and thus we can write equation \eqref{14} in the form:
\begin{equation*}\label{17}
\left( \frac{\partial^2 }{\partial x_{\nu}^2} + 2i \Phi_{\nu} \frac{\partial}{\partial x_{\nu}} - \alpha^2 - \Phi_{\nu}^2 \right) F = -i (\nabla \CON{\Phi}) F .
\tag{17}
\end{equation*} 
Quite the same equation holds for $G$ as well, except that $i$ has to be replaced by $-i$ (or $\Phi$ by $-\Phi$).

   Equation \eqref{17} represents Schr\"odinger's wave equation completed by the spin interaction, which follows from Dirac's theory as a specific effect without any special assumption.

   Now we leave the quaternion formalism and try to write our equations in tensor analytical interpretation where we shall again regard $F$ to be an antisymmetric tensor (invariant in its time part), and $G$ as a complex vector. We use the same notation as in our first discussion [see equations (96)], except that we now prefer a way of description which eliminates the occurrence of imaginary quantities in general. To this end, we leave the four Minkowskian coordinates and regard the real time as being introduced as a fourth coordinate. Then we only have to distinguish between covariant and contravariant. At the same time, however, we can also bring the system into a generally covariant form  in which the system remains invariant not only under linear but under any point transformation. This means that our equation system is related to arbitrary curvilinear coordinates.

   On the one hand, we have the system \eqref{18A} which is analogous to Maxwell's equations and is now modified as follows:
\begin{equation*}\label{18A}
\left.
\begin{aligned}
\frac{\partial S}{\partial x_{\nu}} g^{i \nu} + \frac{1}{\sqrt{g}} \frac{\partial \sqrt{g} F^{i \nu}}{\partial x_{\nu}} & = \alpha S^i - \Phi_{\nu} \DUA{F}^{i \nu} - \Phi^i M , ~~~ ~~~ \\
\frac{\partial M}{\partial x_{\nu}} g^{i \nu} + \frac{1}{\sqrt{g}} \frac{\partial \sqrt{g} \DUA{F}^{i \nu}}{\partial x_{\nu}} & = \alpha M^i + \Phi_{\nu} F^{i \nu} + \Phi^i S . 
\end{aligned}
\right\}
\tag{18A}
\end{equation*}
This is completed by the following ``feedback system:''
\begin{equation*}\label{18B}
\left.
\begin{aligned}
\frac{\partial S_i}{\partial x_k} - \frac{\partial S_k}{\partial x_i} + \DUA{\left( \frac{\partial M_i}{\partial x_k} - \frac{\partial M_k}{\partial x_i} \right)} & = \alpha F_{ik} + ( \Phi_i M_k - \Phi_k M_i )   ~~~ ~~~ \\
 & \quad - \DUA{( \Phi_i S_k - \Phi_k S_i )}  , \\
\frac{1}{\sqrt{g}} \frac{\partial \sqrt{g} S^{\nu}}{\partial x_{\nu}} & = \alpha S - \Phi_{\nu} M^{\nu} , \\
\frac{1}{\sqrt{g}} \frac{\partial \sqrt{g} M^{\nu}}{\partial x_{\nu}} & = \alpha M + \Phi_{\nu} S^{\nu} .
\end{aligned}
\right\} \tag{18B}
\end{equation*}
Here $g_{ik}$ is the metrical tensor and $g$ is the determinant of the metric, or the determinant multiplied by $-1$ because of the negative inertia index of the space-time line elements.

   For an antisymmetric tensor of second-order, the dual assignment should be performed according to the following pattern:
\begin{equation*}\label{19}
\DUA{F}^{12} = \frac{1}{\sqrt{g}} \, F_{34} , \quad \text{etc.}
\tag{19}
\end{equation*}
\begin{equation*}\label{19'}
\DUA{F}_{12} = \sqrt{g} \, F^{34} , \quad \text{etc.}
\tag{19'}
\end{equation*}

   Let us compare the obtained system with that of the previous paper where the vector potential did not occur (Section 10). Then we shall find a number of remarkable differences, which seem to hinder a simple interpretation of the equations to a larger extent than it was expected before. The previous formulation could be considered as a generalization and completion of Maxwell's equations. The generalization consisted of the occurrence of the two scalars $S$ and $M$, which entered the system because it did not contain any internal dependence, and thus the conservation laws for electric and magnetic current could not be deduced as a necessary consequence. The completion was done by attaching new equations which were called ``feedback'' because we considered them as the reaction of the system upon itself. The emerging manifold of quantities could be reduced by putting both the scalars and the magnetic current equal to zero. In this way, we arrived at a still closer connection to the customary form of Maxwell's equations.
Though we then had 16 equations for 10 functions (the field strength and the electric current), no overdetermination occurred because the surplus equations, as consequences of the remaining ones, were satisfied automatically.

   If we try to proceed in an analogous way here as well, we shall see that nothing can be put equal to zero without producing an overdetermination.  For instance, if we put $S$ and $M = 0$, then the last two equations of system \eqref{18B} will be obtained as a consequence of the others only if $\Phi_i$ are constant, otherwise, an additional term will appear which depends on the external field strength and does not vanish.
Similarly, we must not put the ``magnetic current'' equal to zero if we want to avoid an overdetermination.

   However, it is very unlikely that such a manifold of quantities would really appear.  In view of the fact that we have one vector potential only, the appearance of two vectors $S_i$ and $M_i$ cannot be understood. We could expect that by putting one of the vectors equal to zero, we could utilize the resulting overdetermination as field equations for the vector potential. However, it does not seem possible to bring the $\Delta$-equations of the vector potential into such a relationship. Thus, we should take into account the complete system which also contains the four equations for the vector potential, without the three vectors $S_i$, $M_i$, $\Phi_i$ having an essential inner connection with each other. Of course, this appears to be quite unbelievable.

   Furthermore, we have no reference point for understanding the characteristic correction terms which appeared in the equations because of the vector potential.

   Finally, we wish to point out a characteristic difficulty of another type caused by the fact that the dual constructions occur in the equations in combination with the non-dual ones.

   In terms of tensor analysis, the dual forms have the following meaning: In addition to the fundamental metrical tensor $g_{ik}$ which is a symmetric tensor of second-order, there is a second fundamental tensor of $n$th-order (i.e., of fourth-order in the four-dimensional space) in each manifold. This second fundamental tensor has a peculiar structure: it is antisymmetric in all indices. This means that all components vanish where any two of the indices are equal. The non-vanishing components are defined as follows: The covariant components are equal to:
\begin{equation*}\label{20}
\sqrt{g} \,\eta_{iklm} ,
\tag{20}
\end{equation*}
and the contravariants are equal to;
\begin{equation*}\label{20'}
\frac{1}{\sqrt{g}} \,\eta^{iklm} ,
\tag{20'}
\end{equation*}
where $\eta$ means the following: $\eta = +1$ if the permutation $iklm$ of the
four figures 1 to 4 is even, and $\eta = -1$ if this permutation is odd. With the help of this tensor we can construct from four vectors the
invariant:
\begin{equation*}\label{21}
\frac{1}{\sqrt{g}} u_{\mu} v_{\nu} w_{\rho} r_{\sigma} \eta^{\mu \nu \rho \sigma} , 
\tag{21}
\end{equation*}
and we obtain the determinant of the four vectors in the numerator, and therefore we call this tensor a ``determinant tensor.''

   The dual formations of tensor analysis are produced by normal multiplication using just this special tensor. For example, for an antisymmetric tensor of second-order $F_{ik}$, we can construct the following new tensor:
\begin{equation*}\label{22}
\DUA{F}^{i k} = \frac{1}{2} \frac{1}{\sqrt{g}}F_{\mu \nu} \eta^{\mu \nu ik} .
\tag{22}
\end{equation*}
This is just the ``dual'' tensor. In this way, we can obviously find a corresponding antisymmetric dual formation of $(n-m$)th-order for each antisymmetric formation of $m$th-order.

   The determinant tensor is of the same type as any other tensor except that a square root is employed for its definition, whereby the sign becomes undefined. It is easy to see that in the case of a reflection, the determinant tensor obeys the transformation of a common tensor only if the latter changes the sign of the square root. Accordingly, one should determine the sign of $\sqrt{g} $, e.g., by prescribing $+1$ in all ``right-hand'' systems and $-1$ in all ``left-hand'' ones. However it is not possible to characterize a ``right-hand'' system of coordinates on the basis of invariant principles.

   There remains only the possibility of normalizing the sign of $\sqrt{g} $ generally, e.g., to $+1$. Then all formations created with the determinant tensor have the peculiarity that the normal transformation formulae used in their transformation have to include the factor $-1$ if we perform a transformation with negative determinant (reflection) (so-called ``axial'' formations in contrast to the ``polar'' ones).

   To explain these conditions, which may not be generally known in the above connection, let us consider an example from three dimensional tensor analysis. We can write the first system of Maxwell's vacuum equations as follows (time is now treated as a common scalar quantity):

\begin{equation*}\label{23}
\frac{1}{c} \frac{\partial E^i}{\partial t} = \frac{1}{\sqrt{g}} \frac{\partial H_{\nu}}{\partial x_{\mu}} \eta^{\mu \nu i}.
\tag{23}
\end{equation*}
(Here $g$ is the determinant of the spatial line element). Now we can either define the sign of $\sqrt{g} $ as positive or negative, depending on whether our system is a right- or a left-hand one (which, as we mentioned before, would make necessary a non-invariant distinction and therefore it would not correspond to the spirit of the general covariance). Then both $E_i$ and $H_i$ are ordinary vectors and the change of the sign due to a reflection takes place within the equation. The other alternative is to normalize the sign of $\sqrt{g} $ to $+1$. Then either $H_i$ or $E_i$ must be an ``axial'' vector.  As it is known, in the four-dimensional approach this difficulty disappears. There we can, if we want to, avoid the dual formation in general.\footnote{A.\ Einstein 1916, see W. Pauli, Theory of Relativity (Teubner, 1921) p.\ 631.} However, even if we keep it, dual formations will not occur combined with non-dual ones, and thus the undefined sign of $\sqrt{g}$ will not cause a problem.

   However, in our equation system \eqref{18A} and \eqref{18B} this is in fact the case. If we want to keep our invariant standpoint, we are compelled to consider $M_i$ as an ``axial vector'' and $M$ as an ``axial scalar.'' Interestingly enough, in the following we should also treat $\Phi_i$ as an ``axial vector.'' This is, however, out of the question.\footnote{\emph{Editorial note:} Here, Lanczos is dismissing the possibility of ``axial vector'' (i.e., pseudo-vector) particles, which are actually allowed by his fundamental equation \eqref{2}.} For even if we suppose the less probable case where the vector potential would be an axial vector, then the same should apply to the electric current. Since, however, velocity is certainly polar, charge should become an axial scalar. In $\Phi_i$ the charge is now multiplied by the vector potential. This means that the two sign changes disappear here as well, so a polar vector is produced in every case.

   Now there is nothing left to do but to maintain the ambiguity of $\sqrt{g} $ and to distinguish the ``right-hand'' coordinate systems from the ``left-hand'' ones --- which means a certain concession to the general covariance.

   It is interesting to note that this difficulty does not occur in Einstein's new geometry of distant parallelism. Since there the metric is already composed quadratically from the fundamental quantities, the root of the determinant is accordingly replaced by the determinant of the fundamental quantities itself.  The determinant tensor behaves under reflection in the same way as a common tensor, and the distinction between polar and axial quantities becomes insignificant. Therefore the dual formations have a much more natural character in this theory than in Riemannian geometry in which the antisymmetric element in general represents by its very nature something strange.

   We have formally managed to bring the Dirac equation into a form which completely satisfies the demands of customary tensor calculus and which suggests a pure field theoretical description. If the obtained equation system is unsatisfactory at some points and also displays inexplicable elements, this is perhaps not to be interpreted as a proof that the course followed here is misleading. The close relation of Dirac's operator to the Hamiltonian quaternion operator, on the one hand, and the close relation of this operator to the four-dimensional tensor analysis, on the other, suggest that the connection found must be more than a superficial coincidence. The field theoretical viewpoint chosen here, which only permits operations with a tensor analytical meaning and requires an interpretation for imaginary quantities in terms of real quantities as well,\footnote{In the present description, it is after all the tensor analytical operation of the dual formation which corresponds to $i$.} may lead to a heuristic approach for a natural improvement of Dirac's theory, by which its conversion into tensor analytic form may provide an equation system which displays more internal consistency and has a greater probability of being proven correct than the one obtained here.

Berlin-Nikolassee, July 1929.

\end{document}